\documentclass[structabstract]{aa}  
\usepackage{graphicx}
\usepackage{epstopdf}
\usepackage{color}

\usepackage{txfonts}
\usepackage{natbib}
\def\gr{$\gamma$-ray}
\begin{document}

\title{Very-high-energy gamma-ray emission from high-redshift blazars}

    \author{A. Neronov
          \inst{1},
          D.Semikoz\inst{2,3}, {A.M.Taylor} \inst{1}
          \and
          Ie.Vovk
          \inst{1}
          }

   \institute{ISDC Data Centre for Astrophysics, Ch. d'Ecogia 16, 1290, Versoix, Switzerland \\
              \email{Andrii.Neronov@unige.ch}
         \and
             APC, 10 rue Alice Domon et Leonie Duquet, F-75205 Paris Cedex 13, France \and
Institute for Nuclear Research RAS, 60th October Anniversary prosp. 7a, Moscow, 117312, Russia
\\
             \email{Dmitri Semikoz <dmitri.semikoz@apc.univ-paris7.fr>}
             }

\abstract
{}
{We study the possible detection of and properties of very high-energy (VHE) \gr\ emission (in the energy band above 100~GeV) from high redshift sources.}
{
We report on the detection of VHE \gr\ flux from blazars with redshifts $z>0.5$. We use the data of Fermi telescope in the energy band above 100~GeV and identify significant sources via cross-correlation of arrival directions of individual VHE \gr s with the positions of known Fermi sources.  }
{There are thirteen high-redshift sources detected in the VHE band by Fermi/LAT telescope. The present statistics of the Fermi signal from these sources is too low for a sensible study of the effects of suppression of the VHE flux by pair production through interactions with Extragalactic Background Light photons. We find that the detection of these sources with ground-based \gr\ telescopes would be challenging. However, several sources including BL Lacs PKS 0426-380 at $z=1.11$,  KUV 00311-1938 at $z=0.61$,  B3 1307+433 at $z=0.69$, PG 1246+586 at $z=0.84$, Ton 116 at $z=1.065$  as well as a flat-spectrum radio quasar 4C +55.17 at $z=0.89$ should be detectable by HESS-II, MAGIC-II and CTA. A high-statistics study of a much larger number of VHE \gr\ sources at cosmological distances would be possible with the proposed high-altitude Cherenkov telescope 5@5.  
}
{}

\maketitle

\section{Introduction}

\gr\ emission from distant blazars is suppressed by the effect of pair production through interactions of these \gr s with the low-energy photons forming the Extragalactic Background Light (EBL) \citep{gould67,kneiske,stecker06,mazin07,franceschini08,gilmore}. This prevents observations of high-redshift sources using the technique of imaging the Cherenkov emission produced through \gr\ induced air showers, used by the ground-based Cherenkov telescopes HESS, MAGIC and VERITAS \citep{aharonian_review}. Most of the VHE \gr\ loud blazars, detected by these telescopes, are situated in the local Universe, at (several) hundred megaparsec distances, in the redshift range $z\sim 0-0.2$\footnote{\tt http://tevcat.uchicago.edu}. Only one source at redshift $z>0.5$, 3C 279, was possibly detected by the MAGIC telescope during a flaring activity \citep{3C279_MAGIC}. Another source at redshift 0.444, 3C 66A, was detected by VERITAS \citep{3c66a}. One more relatively high redshift source, PG 1553+113 (at $z>0.4$, reported by  \citet{danforth}) has been detected by MAGIC and VERITAS \citep{PG1553_MAGIC,PG1553_Veritas}.

Measurement of the effect of suppression of the VHE \gr\ flux from low-redshift blazars is commonly used for the estimation of the density of the EBL in the local Universe \citep{kneiske,stecker06,mazin07,franceschini08,gilmore}. Constraints on the EBL density and spectral characteristics, extracted from the VHE blazar observations, are useful for understanding the cosmological evolution of galaxies. Observations of the suppressed VHE \gr\ emission from blazars at non-negligible redshifts can provide a measurement or constraint not only of the EBL density in the local Universe, but also new information on the cosmological evolution of the EBL, which is largely uncertain.   

Observations of high-redshift sources in the VHE band would also provide valuable information on the cosmological evolution of the VHE \gr\ loud blazars (and, more generally, \gr\ emitting active galactic nuclei, AGN), which is also highly uncertain. This information is important, in particular, for an understanding of the origin of Extragalactic \gr\ background  \citep{fermi_background,neronov_EGB}. Most of the VHE \gr\ emitting blazars at zero redshift belong to a High-energy peaked BL Lac (HBL) sub-class of the blazar population. Negative cosmological evolution of this sub-class was reported based on the observations in the visible and X-ray band \citep{giommi}. Such puzzling evolution is, apparently, opposite to the general positive evolution of blazars and radio galaxies (BL Lac parent AGN population) with the cumulative power of the sources increasing as $(1+z)^k$, $k>0$ \citep{hodge09,sadler07,smolcic09}. Independent verification of this hypothesis using \gr\ observations would be possible only if a significant amount of VHE emitting blazars could be detected in the redshift range $z\sim 0.5-1$.  

Taking into account the importance of the study of the VHE \gr\ emission from distant AGN, we report here on the detection of VHE \gr\ signals from sources at redshifts $z>0.5$ by the Large Area Telescope (LAT) on board of Fermi satellite \citep{atwood09}. The effective area of LAT, about 1~m$^2$, is several orders of magnitude smaller than that of the ground-based \gr\ telescopes, so that LAT has detected only one-to-several photons from the brightest extragalactic VHE \gr\ sources in $\sim 4$~years of observations. However, an extremely low level of the background (including the residual cosmic ray background and Galactic / extragalactic \gr\ background) makes even a several-photon signal in the energy band above 100~GeV significant. This fact has been used by \citet{ic310,100GeV_paper1} for the search of extragalactic VHE \gr\ sources via the analysis of clustering of VHE \gr s on the sky, based on the method first proposed by \citet{Gorbunov_method}. In this paper we use the correlation of arrival directions of VHE \gr s detected by Fermi with the positions of high-redshift blazars to identify high-redshift sources of VHE \gr s and to study their spectral characteristics.

\section{Data selection and data analysis}

For our analysis we have used Fermi pass 7 data within the time window from August 4, 2008 up to June 11, 2012. We have considered two sub-classes of the "ULTRACLEAN" events ("evclass=4" in the photon selection routines): the event from the classes 65311 and 32543.   Those classes have the best angular resolution in Fermi, corresponding to a "superclean class"  in pass 6 data for $E>100$~GeV.  
Using the method of \citet{Gorbunov_method} we correlated the arrival directions of photons with $E>100$~GeV 
with positions of the sources listed in the two-year Fermi catalog \citep{fermi_catalog} with high Galactic latitude $|b|>10^\circ$. This analysis is similar to one done for the 1FGL catalog by \citet{100GeV_paper1}, where a catalog of 75~objects with $E>100$~GeV was obtained.

There are  1036 front and 888 back converted photons  at high Galactic latitude $|b|>10^\circ$ and $ E>100$~GeV in  the selected data.

We have measured the PSF for the selected events via the correlation of photon arrival directions with the sky positions of known gamma-ray blazars.  The size of the  68\% containment circle for  the event classes 65311 and 32543 is  $0.08^\circ$ for front photons and $0.21^\circ$ for back photons with energies above 100~GeV.
In order to keep the analysis methods consistent with those used by {\citet{100GeV_paper1}, we use  $0.1^\circ$ and $0.2^\circ$  distance bins for correlation analysis using front and back photons.

We took all high Galactic latitude Fermi LAT sources with known redshifts and selected high redshift sources with $z>0.5$. The original two-year Fermi catalog \citep{fermi_catalog} does not contain information on the redshift of the sources. Instead, we have considered the redshifts listed in the 13th Veron catalog of 
 BL Lacs~\citep{veron13} and in  the Rome blazar  catalog of \citet{FSRQ}. We cross-checked the information on the source redshifts with the Fermi AGN catalog \citep{fermi_agn} and with the information given in the NED\footnote{\tt http://ned.ipac.caltech.edu/} and SIMBAD\footnote{\tt http://simbad.u-strasbg.fr/simbad/} databases. In this way out of 414 Fermi BL Lacs we selected 55 with $z>0.5$. 

For the 55 selected BL Lacs we found six  photons correlating with six objects within 0.1 degree, while a total background of 
0.06 was expected. The probability that this signal occurred by chance was $P  \sim 6 \cdot 10^{-11}$.  The six sources contributing to the correlation are listed in the upper part of the Table \ref{tab:list}. Within 0.2 degrees from the source positions, (but outside the 0.1 degree region of the sources) 
the expected background was 0.14 photons, while 4 photons were observed from 3 additional BL Lacs. 

One of these three additional sources was KUV 00311-1938, with two associated photons. The chance coincidence probability for the two photons to arrive within 0.2 degrees from the source position was $8\times 10^{-6}$, which corresponds to a $4.5\sigma$ detection of the source above 100~GeV. The chance coincidence probability for the remaining two photons to arrive within $0.2^\circ$ from two other BL Lacs was $P \sim 10^{-2}$.
We list these sources in the lower part of the Table \ref{tab:list} (two last lines), but one should remember that they might, in principle, be "fake detections" due to the chance coincidence of the background photon arrival directions.

From the 351 Flat Spectrum Radio Quasars (FSRQ) we selected  269 objects with $z>0.5$. Within 0.1 degree from these objects we found four photons from four different sources, 
while a total background of 0.4 was expected. The probability that this happened by chance was  $P \sim 8 \cdot 10^{-4}$. Again, these objects were given lower 
priority in Table 1, similar to that for the case of BL Lacs with photons only associated within 0.2 degrees.
We did not consider correlations with FRSQ objects within larger angles because of the high level of chance coincident background.

The complete list of sources with redshift $z>0.5$ which correlate with the arrival directions of the VHE \gr s in LAT is shown in Table \ref{tab:list}. 

For the spectral and timing analysis presented in Figs. 1-13, we have used the Fermi Science Tools \footnote{http://fermi.gsfc.nasa.gov/ssc/data/analysis/}. We have calculated the source spectra using two complementary techniques: the likelihood analysis and the aperture photometry methods. The two methods give consistent results. The likelihood method being more reliable at low energies ($0.1-10$~GeV) where the LAT instrument has a relatively wide point spread function. The aperture photometry method enables to properly take into account the Poissonian statistics of the signal at low photon count rates in the 10-300~GeV range.

\section{Properties of individual high-redshift sources}

\begin{table*}
\begin{tabular}{|c|c|c|c|c|c|c|c|c|c|c|c|}
\hline
&Name&RA&DEC&Type&$z$&$N_{30-100}$ &$N_{0.1}$ &$N_{0.2}$&$E_{max}$&$L/L_{\rm Mrk 421}$& $P$\\
\hline
1&TXS 0138-097  &25.3576&-9.4788& BL   &0.733  & 2   &1f   & 0   &138&44.5  &1.1e-3 \\
2&PKS 0426-380  &67.1685&-37.9388& BL   &1.11    & 13 &1f   & 0   &134&151  &0.9e-3  \\
3&B2 0912+29      &138.9683&29.5567& BL   &1.521  & 5   &1f    &0   &126&405  &0.7e-3  \\
4& Ton  116           &190.8031&36.4622&BL    &1.065  & 11 &1b     &0  &114&133  &0.7e-3 \\ 
5& PG 1246+586  &192.0783&58.3413&BL    &0.847  & 9   &1b     &0  &104&67.5  &1.2e-3 \\
6& B3 1307+433   &197.3563 &43.0849&BL    &0.69    & 4   &1f     &0  &104&37.5  &0.8e-3\\   
7&4C +55.17&149.4091&55.3827&FSRQ &0.8955& 14  &1b &0           &141&  84.0      &1.6e-3\\       
8&TXS 1720+102&260.6857&10.2266&FSRQ&0.732& 0 &1f&0        &168&  46.8           &1.9e-3 \\
9& PKS 1958-179&300.2379&-17.8160&FSRQ&0.65 &2  &1b &0        &118& 33.5            &2.3e-3  \\
10&PKS 2142-75&326.8030&-75.6037&FSRQ&  1.139 &1  &1f &0       & 135&173            &1.5e-3   \\
\hline
11 & KUV 00311-1938 &8.3933&-19.3594& BL   & 0.61 &11& 0 &2b  &152 &53.2 &8e-6  \\
12&RGB J0250+172  &42.6567&17.2067&BL    &1.1    &3   &0  &1b  &358 &147 &7.6e-3 \\
13&PKS 1130+008     &173.190067&0.5744&BL    &1.223&1  & 0 &1f  &140 &204 &4.4e-3   \\
\hline
\end{tabular}
\caption{List of high-redshift blazars emitting in the energy band $E>100$~GeV.  Columns $N_{0.1}$ and $N_{0.2}$ give the numbers of VHE photons within an angular distance of $0.1^\circ$ and $0.2^\circ$ from the source position. $P$ is the chance coincidence probability for the VHE photons to be found within the $0.1^\circ$ or $0.2^\circ$ circles from the source labels "f" and "b" mark photons pair converted in the front- and back- layers of the LAT detector.  Column $N_{30-100}$ gives the number of photons associated to the source in the lower energy band $30-100$~GeV. $E_{max}$ is the maximal energy of photon associated to the source. $L/L_{Mrk\ 421}$ is the source luminosity in units of Mrk~421 luminosity at $E>100$~GeV.}
\label{tab:list}
\end{table*}

In this section we study in more detail the spectral and timing properties of the high-redshift VHE \gr\ sources listed in Table \ref{tab:list}, to understand if some of the lower significance sources in Table \ref{tab:list} among the FSRQs and the two BL Lacs listed in the last two rows of the table could possibly be random coincidences of arrival directions of the VHE \gr s with the source positions on the sky. We also investigate the possibility for the detection of these sources using ground-based \gr\ telescopes.

\subsection{Firm detections}

\subsubsection{TXS 0138-097}

\begin{figure}
\includegraphics[width=\linewidth]{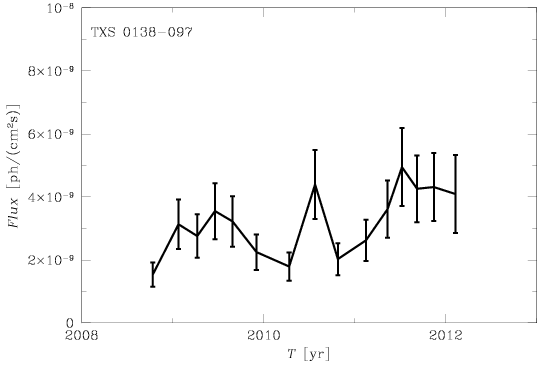}
\includegraphics[width=\linewidth]{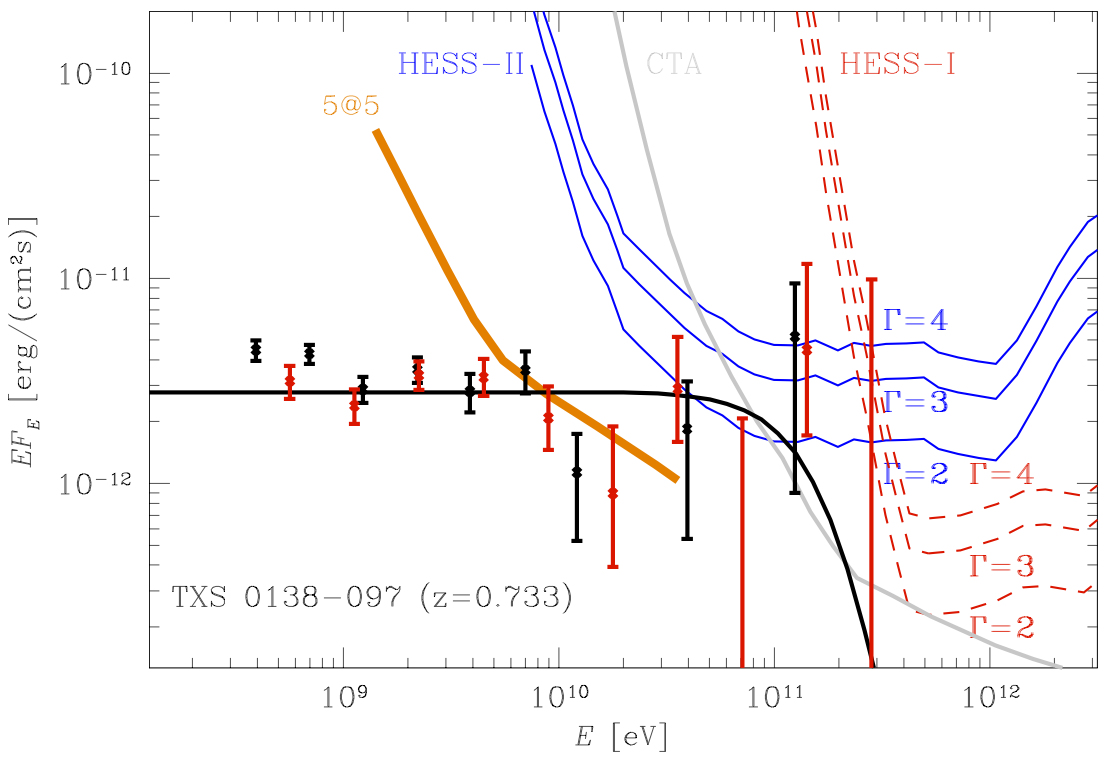}
\caption{Lightcurve (top) and spectrum (bottom) of TXS 0138-097 at redshift $z=0.733$.  Black curve in the bottom panel shows a powerlaw type model spectrum absorbed on the EBL from \citet{franceschini08}.  Also shown in the bottom panel are sensitivity curves of different ground based \gr\ telescopes: 5@5 \citep{5at5}, HESS-I and HESS-II \citep{masbou} and CTA \citep{CTA}. Black and red data points in the lower panel show the spectrum extracted using the likelihood analysis and aperture photometry method, respectively.}
\label{fig:0138}
\end{figure}

The redshift of this source has been obtained through  spectroscopic measurements \citep{stocke97}. At the same time, the SIMBAD database entry for the source gives a redshift of 1.03, based on the data from the SDSS photometric catalog \citep{sdss_0138}. The detection of absorption lines at redshift $z=0.5$ \citep{stickel} for this source imposes a lower limit on its redshift, so that in any case the source belongs to the $z>0.5$ source sample in which we are interested in this paper. 

The upper panel of Fig. \ref{fig:0138} shows the lightcurve of the source in the energy band above 1~GeV. From this lightcurve, one can see that the source behaviour is stable on the time scale of almost four years of Fermi exposure. The lower panel of the same figure shows the source spectrum in the 0.3-300~GeV energy range. The spectrum is well described by a simple power law of the form $dN/dE\sim E^{-\Gamma}$ with spectral slope $\Gamma\simeq 2$, suppressed at high-energies through absorption on the EBL. The model spectrum shown in Fig. \ref{fig:0138} is calculated assuming a source redshift of 0.733 and using the EBL model of  \citet{franceschini08}. The 100-200~GeV band source flux is consistent with the expectation based on the Franceschini et al. model. 

Of course, the statistics of the Fermi signal is largely insufficient for a sensible study of the details of the shape of the spectrum in the 100~GeV band, which is determined by the effects of propagation through the EBL. Such a study would be possible only with a ground-based \gr\ telescope, which provides a much larger collection area for \gr s and, consequently, much higher source signal statistics. The lower panel of Fig.~\ref{fig:0138} shows a comparison of the measured source spectrum with the sensitivity limits of different existing and future ground-based \gr\ telescopes. Unfortunately, the model source flux is below the sensitivity limit of the  HESS-II telescope, which will start operation in 2012. The estimate of the HESS-II sensitivity shown in this plot has been taken from \citet{masbou}. The sensitivity curve found in \citet{masbou} is given in terms of the integral flux. Conversion of the sensitivity into the differential flux sensitivity depends on the slope of the source spectrum. Fig. \ref{fig:0138} shows three different differential sensitivity curves corresponding to three different values of the slope of the source spectrum. One can see that within the energy range covered by HESS-II, the spectrum is steep, with a slope close to $\Gamma=4$, so that the source flux is below the HESS-II sensitivity level such that the source is not expected to be detectable. 

The expected performance of the next-generation ground-based \gr\ telescope CTA is not better than that of HESS-II for energies below 100~GeV. Because of this, the prospects for the source detection with CTA are also not promising. Contrary to this, however, the sensitivity of a ground-based \gr\ telescope optimized for the 10~GeV energy band, 5@5 \citep{5at5} is sufficient for the source detection in the energy range above 10~GeV and up to the sharp EBL-induced cut-off at 100~GeV. The difference in the performance of 5@5 and CTA in the 10-100~GeV energy band turns out to be crucially important for the possibility of the study of VHE \gr\ emission from  this high-redshift source.

\subsubsection{PKS 0426-380}

\begin{figure}
\includegraphics[width=\linewidth]{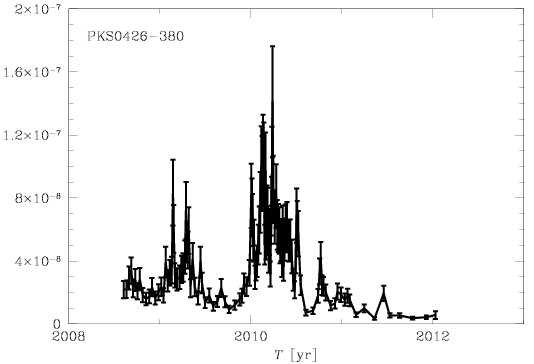}
\includegraphics[width=\linewidth]{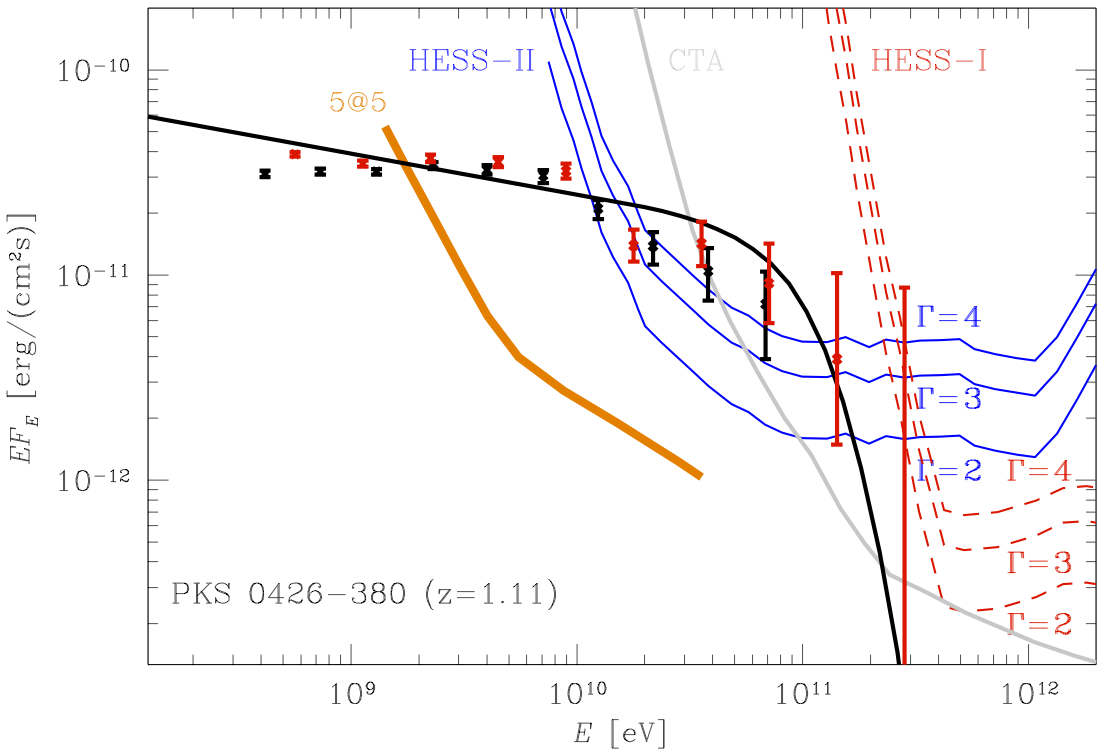}
\caption{Lightcurve (top) and spectrum (bottom) of PKS0426-380 at the redshift $z=1.11$.  Notations are the same as in Fig. \ref{fig:0138}. }
\label{fig:0426}
\end{figure}

This bright BL Lac is situated at a redshift of $z=1.11$ \citep{0426_redshift}. The source is strongly variable in the GeV energy band, see upper panel of Fig. \ref{fig:0426}. Several active episodes, during which the flux increased by up to an order of magnitude, have been observed during the four years of Fermi's operation. 

The source spectrum averaged over the whole observation period extends as a soft powerlaw with $\Gamma>2$ well into the 100~GeV energy band at the level of $\sim 10^{-11}$~erg/cm$^2$s, see lower panel of Fig. \ref{fig:0426}. At this flux level, the source should be readily detectable in the 30-100~GeV energy range with the HESS-II telescope and, in the future, with CTA. 

We have verified that the source flux did increase by a factor of $\sim 2-3$ during the flaring activities, without any significant change in its spectral shape. This implies that the detectability of the source from the ground should be facilitated if the next ground-based observations are triggered during the next increased activity episode (which could be traced based on the Fermi monitoring of the source). 

\subsubsection{B2 0912+29}

\begin{figure}
\includegraphics[width=\linewidth]{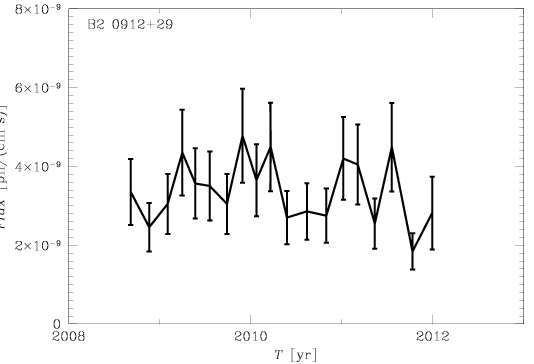}
\includegraphics[width=\linewidth]{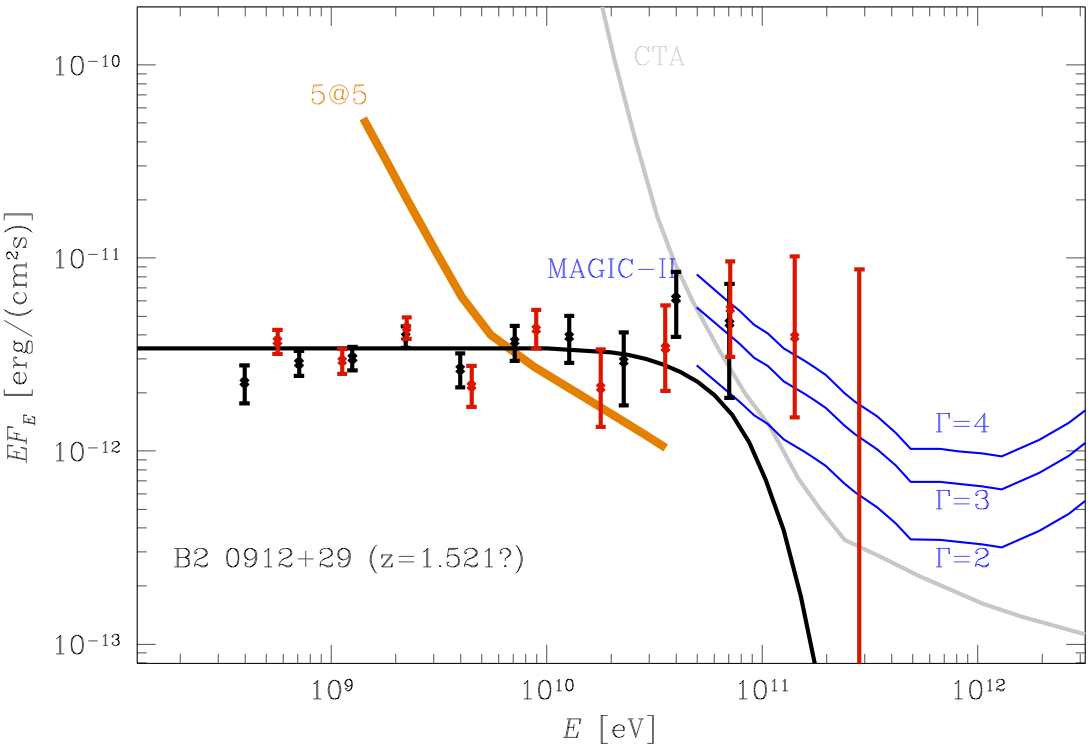}
\caption{Lightcurve (top) and spectrum of B20912+29 at the supposed redshift $z=1.521$.  Notations are the same as in Fig. \ref{fig:0138}. Blue curves show the sensitivity of the MAGIC telescope in the stereo observation mode, from {\citet{MAGIC_sensitivity}}.}
\label{fig:0912}
\end{figure}

The SIMBAD database ascribes a redshift $z=1.521$ to the source, with a reference to the SDSS photometric survey \citep{sdss_0138}, but stating that the redshift is spectroscopic. At the same time, the NED database does not ascribe a redshift to the source. 

The source spectrum can be described by a relatively hard powerlaw with slope $\Gamma\simeq 2$ up to $\sim 200$~GeV energy, showing no sign of suppression by the EBL. Such an unabsorbed powerlaw shape of the spectrum would pose a problem for modeling of the source if it is indeed at a redshift of $z=1.5$. At this redshift, the suppression of the flux due to absorption on the EBL should start already below 100~GeV, so that the flux measurements in both 50-100~GeV and 100-200~GeV bands would be in tension with the model, see Fig. \ref{fig:0912}.

The bottom panel of Fig. \ref{fig:0912} shows a comparison of the sensitivity of the MAGIC telescope in stereo operation mode \citep{MAGIC_sensitivity} {with the level of the source flux for different assumed values of the slope of the powerlaw spectrum $\Gamma$. One can see that if the source flux is indeed at the level measured by Fermi, rather than the theoretically expected flux from a source at the redshift 1.521 suppressed by the absorption on EBL, this source should be detectable by MAGIC.

\subsubsection{Ton 116} 

\begin{figure}
\includegraphics[width=\linewidth]{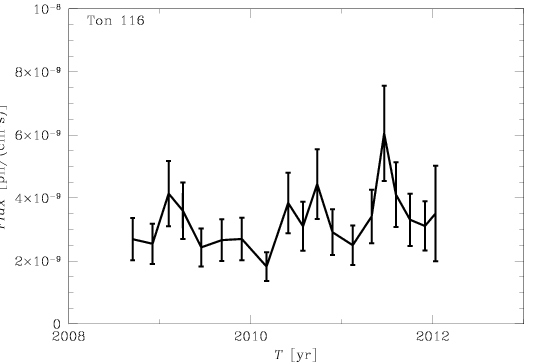}
\includegraphics[width=\linewidth]{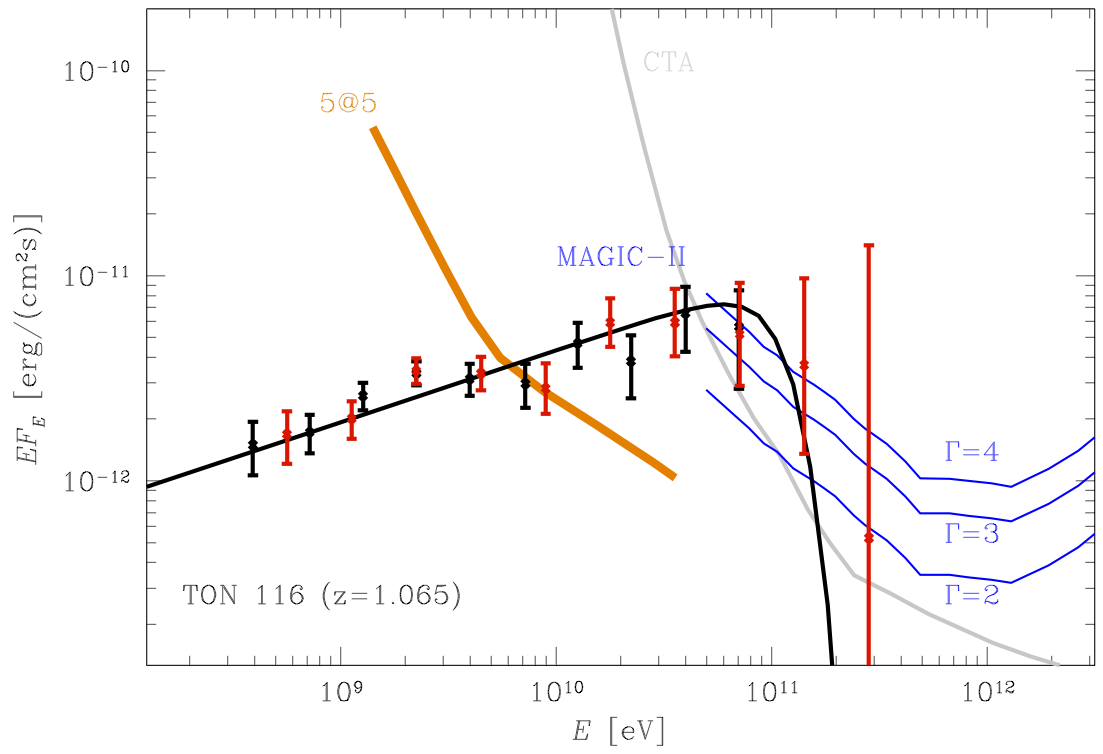}
\caption{Lightcurve (top) and spectrum of TON 116 at the redshift $z=1.065$.  Notations are the same as in Figs. \ref{fig:0138} and \ref{fig:0912}.}
\label{fig:ton116}
\end{figure}

The redshift of the source cited in the NED database is $z=1.065$, with a reference to the SDSS measurement \citep{sdss_release}. At the same time, the Fermi AGN catalog \citep{fermi_agn} list this source as having an unknown redshift. The source exhibits a hard spectrum with slope $\Gamma\simeq 1.7$ (see Fig. \ref{fig:ton116}), so that in spite of the high redshift of the source, the flux in the 100~GeV energy range should be at the level detectable by MAGIC and, in the future, by CTA. The source spectrum is expected to become very steep just above 100~GeV, so that its signal in MAGIC and CTA would be dominated by the contribution from the lowest energy bin just at the threshold energy of the instrument. It is not clear what quality of measurement of the shape of the spectrum in the VHE band could be achieved for such a situation. A proper study of the shape of the spectrum and of the effect of the EBL on it could be done only with a dedicated instrument with a very low energy threshold, like 5@5. From Fig. \ref{fig:ton116} one can see that the 5@5 instrument would be able to detect the source starting from an energy of $\simeq 10$~GeV.

\subsubsection{PG 1246+586}

\begin{figure}
\includegraphics[width=\linewidth]{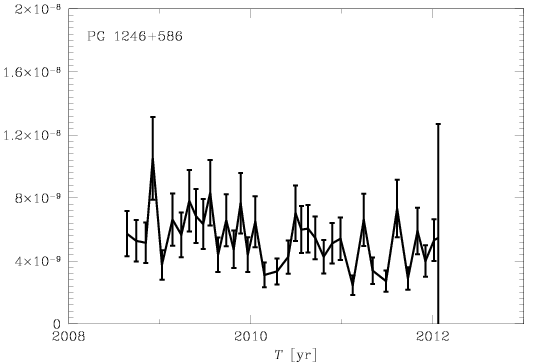}
\includegraphics[width=\linewidth]{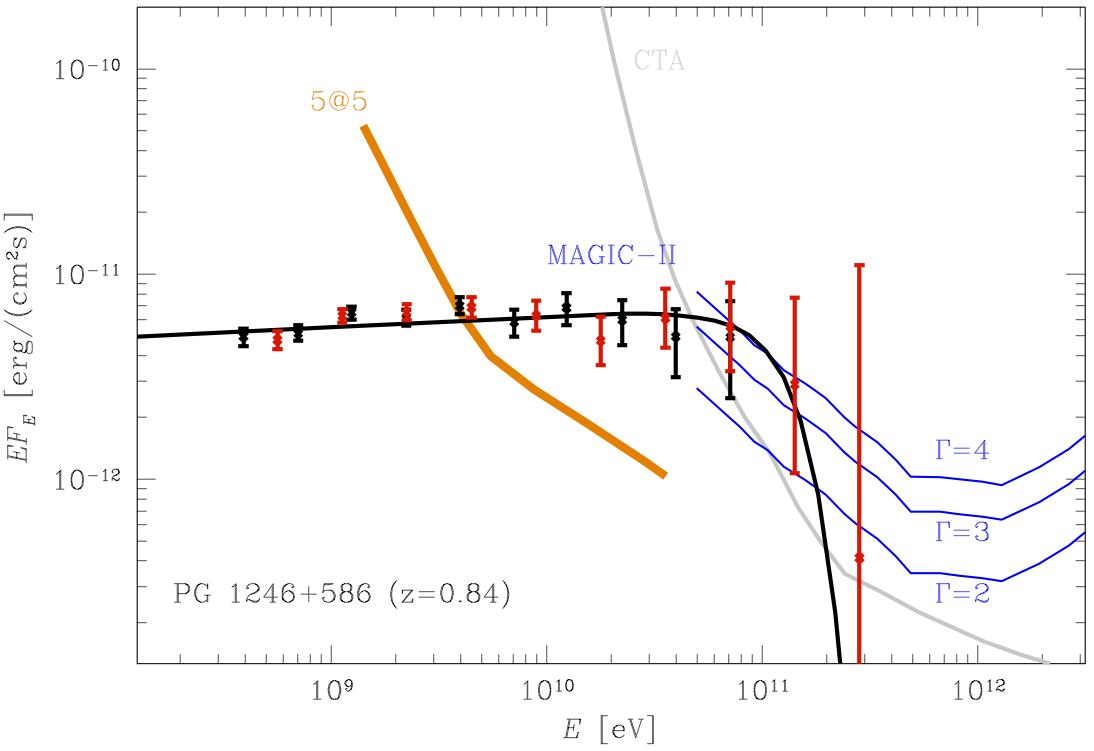}
\caption{Lightcurve (top) and spectrum of PG 1246+586 at the redshift $z=0.847$.  Notations are the same as in Figs. \ref{fig:0138} and \ref{fig:0912}.}
\label{fig:1246}
\end{figure}

The source redshift listed in both the NED and SIMBAD databases is $z=0.847$ with a reference to SDSS \citep{sdss_release}. In the Fermi AGN catalog the source redshift is considered as unknown \citep{fermi_agn}. The source exhibits a steady behavior, remaining stable over the four-years of Fermi monitoring (see the top panel of Fig. \ref{fig:1246}) and the source spectrum is hard with a slope slightly harder than $\Gamma=2$ up to energies above 100~GeV, where the spectrum starts to be affected by the effect of suppression on the EBL (bottom panel of Fig. \ref{fig:1246}). The sensitivity of the MAGIC telescope would be sufficient for the detection of the source, as would the sensitivity of CTA. Similar to the case of Ton 116, a proper study of the shape of the VHE \gr\ spectrum could be done only with a dedicated low energy threshold instrument, like 5@5, which would be able to detect the source starting from an energy $\sim 10$~GeV. 

\subsubsection{B3 1307+433}

\begin{figure}
\includegraphics[width=\linewidth]{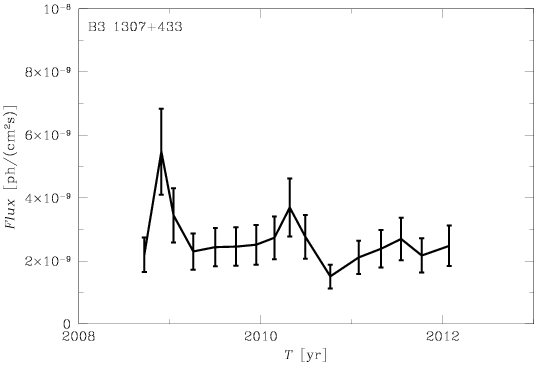}
\includegraphics[width=\linewidth]{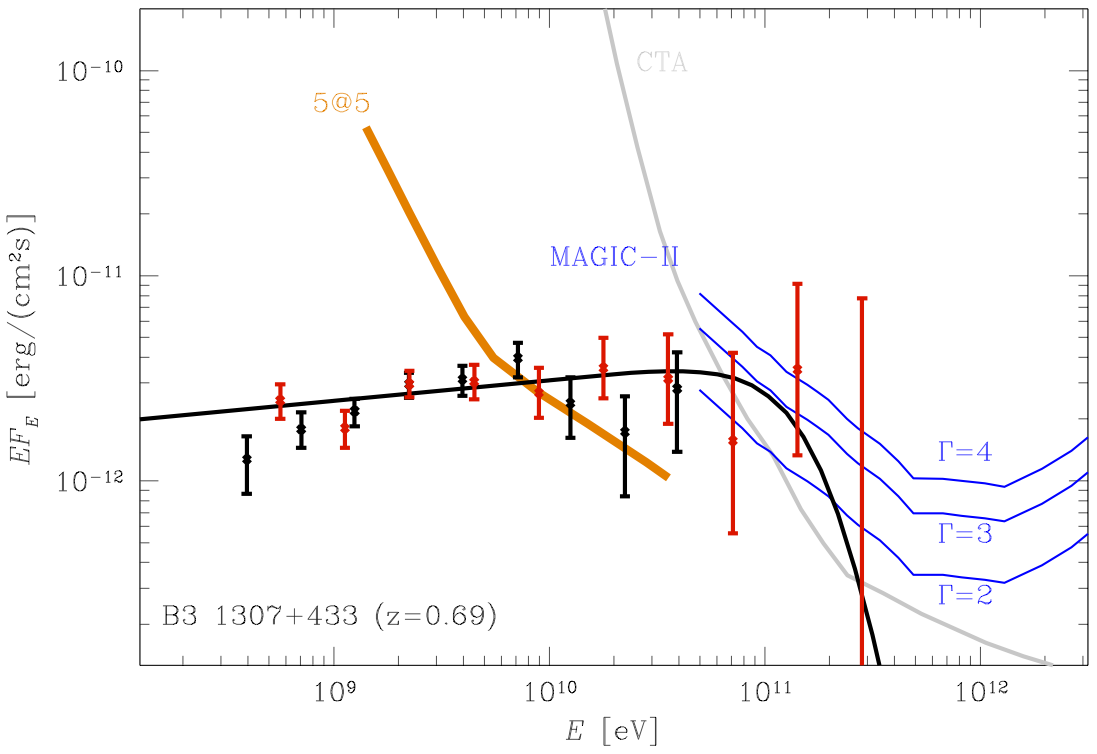}
\caption{Lightcurve (top) and spectrum (bottom) of KUV 00311 at the redshift $z=0.69$. Notations are the same as in Figs. \ref{fig:0138} and \ref{fig:0912}.  }
\label{fig:1307}
\end{figure}

The redshift  quoted for this source in the NED database is $z=0.69$, based on the SDSS measurement \citep{plotkin}. At the same time, SIMBAD quotes a redshift of $z=2.159$, also based on the SDSS data \citep{sdss_0138}. The spectrum of the source,  shown in Fig. \ref{fig:1307}, is a hard ($\Gamma<2$) powerlaw up to $\sim 200$~GeV. 
 If the source were at a redshift of 2.159, as mentioned by SIMBAD, the detection above 100~GeV would be in contradiction with the expected level of suppression of the source flux due to the interactions of \gr s with the EBL photons. At a redshift of 0.69, however, the Fermi measurement is consistent with the theoretical expectation (see Fig. \ref{fig:1307}). 
 
 Verification of the source spectral properties in the 100~GeV band, therefore, would allow clarification of the issue of the source redshift via observations with the existing ground based telescopes (MAGIC and Veritas). However, high quality measurements of the shape of the spectrum would require a next generation instrument, like CTA and 5@5. 
 
\subsubsection{KUV 00311-1938}

As is mentioned in the previous section, this source is individually detected above 100~GeV with significance $4.5\sigma$. The redshift of this BL Lac is $z=0.61$ \citep{giommi05}. Its spectrum is characterized by a hard powerlaw type with $\Gamma <2$ from 0.3 up to $\sim 200$~GeV. The VHE data point in the 100-200~GeV band matches exactly that expected from an extrapolation of the low-energy spectrum, with no signature of the EBL suppression noticable at this energy at the Fermi data quality level. The Fermi measurement, however, is consistent with the assumption of a powerlaw suppressed by the absorption on EBL. The EBL effect is expected to appear mostly above 200~GeV, in the energy band readily accessible by the HESS telescope. The source flux above 100~GeV is sufficiently high for the detection with HESS-II and, possibly, HESS (see Fig. \ref{fig:KUV}).  
  
\begin{figure}
\includegraphics[width=\linewidth]{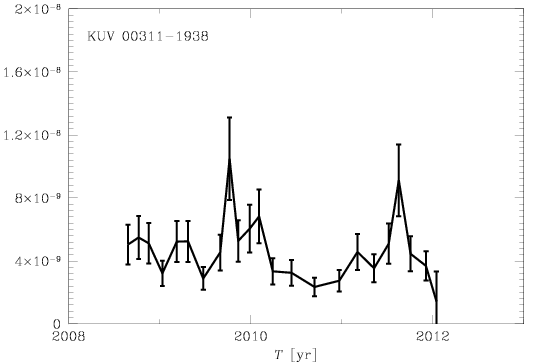}
\includegraphics[width=\linewidth]{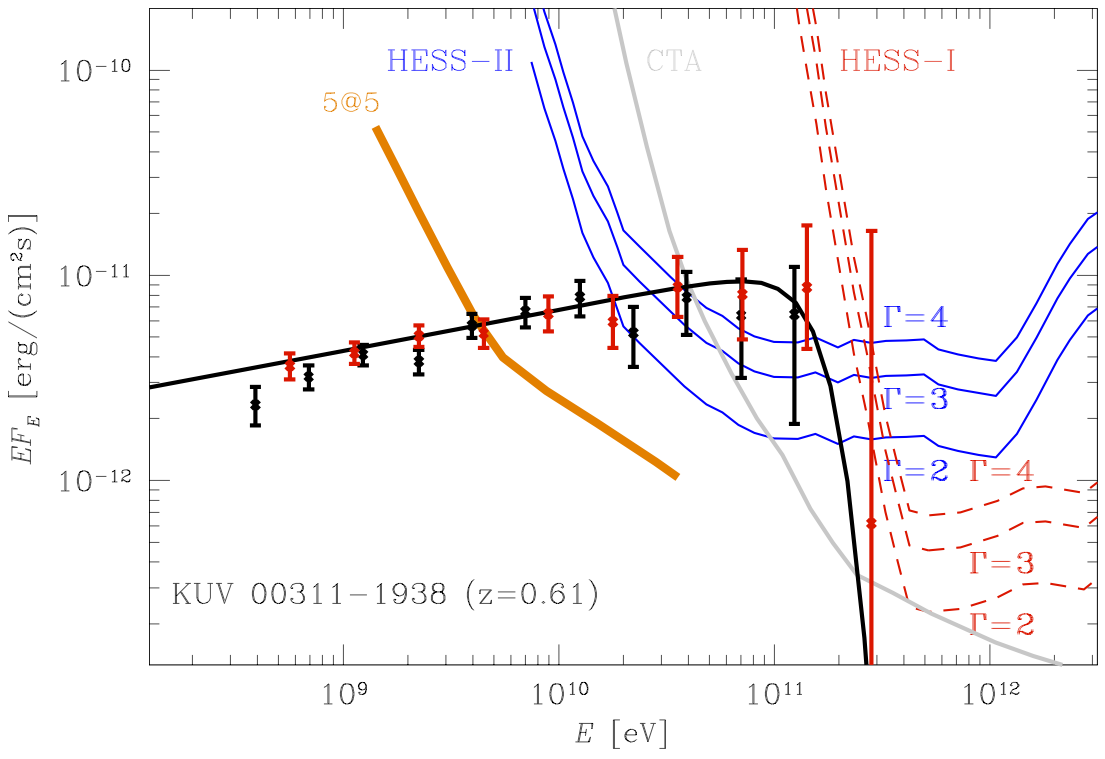}
\caption{Lightcurve (top) and spectrum of KUV 00311-1938 at the redshift $z=0.61$. Notations are the same as in Fig. \ref{fig:0138}.   }
\label{fig:KUV}
\end{figure}

\subsection{Possible detections} 
 
In this subsection we provide the details on sources for which the correlation of the arrival directions of $E>100$~GeV photons with their source positions is less significant. We notice that some "chance coincidence" sources can possibly be singled out based on their spectral characteristics (the extrapolation of the low-energy spectrum to the VHE band is not consistent with the source flux level in the VHE band implied by the detection of the VHE photon). However, this criterium alone can not be used to firmly reject the source due to the possible existence of peculiar source spectra (e.g. due to the presence of different emission components in different energy bands).
 
\subsubsection{4C +55.17}

\begin{figure}
\includegraphics[width=\linewidth]{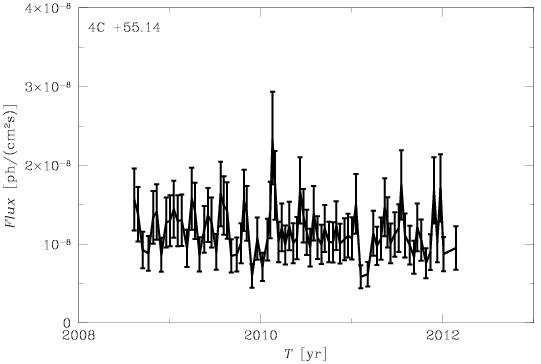}
\includegraphics[width=\linewidth]{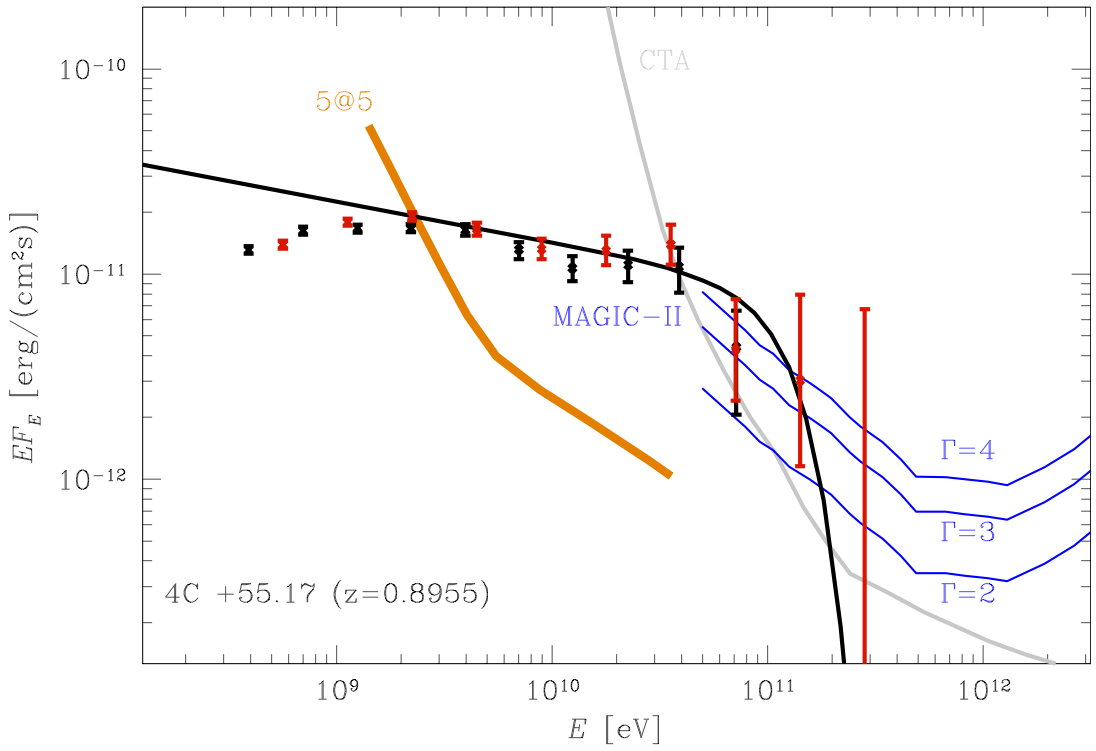}
\caption{Lightcurve (top) and spectrum (bottom) of 4C +55.17at the redshift $z=0.8955$.   Notations are the same as in Figs. \ref{fig:0138} and \ref{fig:0912}.   }
\label{fig:4c55}
\end{figure}

4C +55.17 is a source from the FSRQ subclass of blazars, i.e. from the sub-class which gives only a minor contribution to the set of known VHE \gr\ sources in the local Universe. The appearance of this source, along with the three other FSRQs in the $z>0.5$ VHE source sample is perhaps important since it may provide an indication of the higher FSRQ contribution to the VHE \gr\ source population in the cosmological past. 

The source's redshift is $z=0.8955$ \citep{sdss_4c55}. The source spectrum is softer than $\Gamma=2$ for GeV energies and above, as expected from FSRQs which have on average softer spectra than BL Lacs (see Fig. \ref{fig:4c55}). An extrapolation of the $E<100$~GeV band spectrum into the VHE band is consistent with the estimate of the VHE flux of the source based on the detected VHE photon. This indicates that the VHE \gr\ emission from the source is real. However, detection of the source from the ground  would be a challenging task for the existing ground-based telescopes like MAGIC (see Fig. \ref{fig:4c55}).  At the same time, the source should be readily detectable by the next-generation instruments CTA and 5@5 (Fig. \ref{fig:4c55}).

\subsubsection{TXS 1720+102}

\begin{figure}
\includegraphics[width=\linewidth]{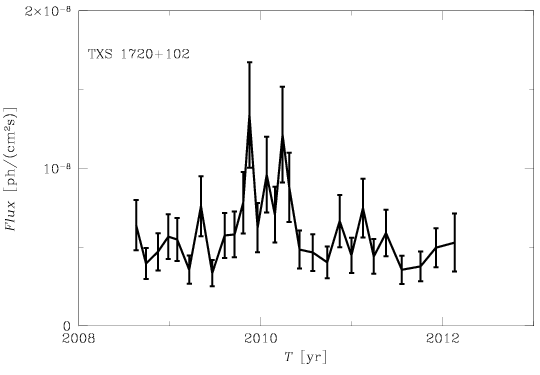}
\includegraphics[width=\linewidth]{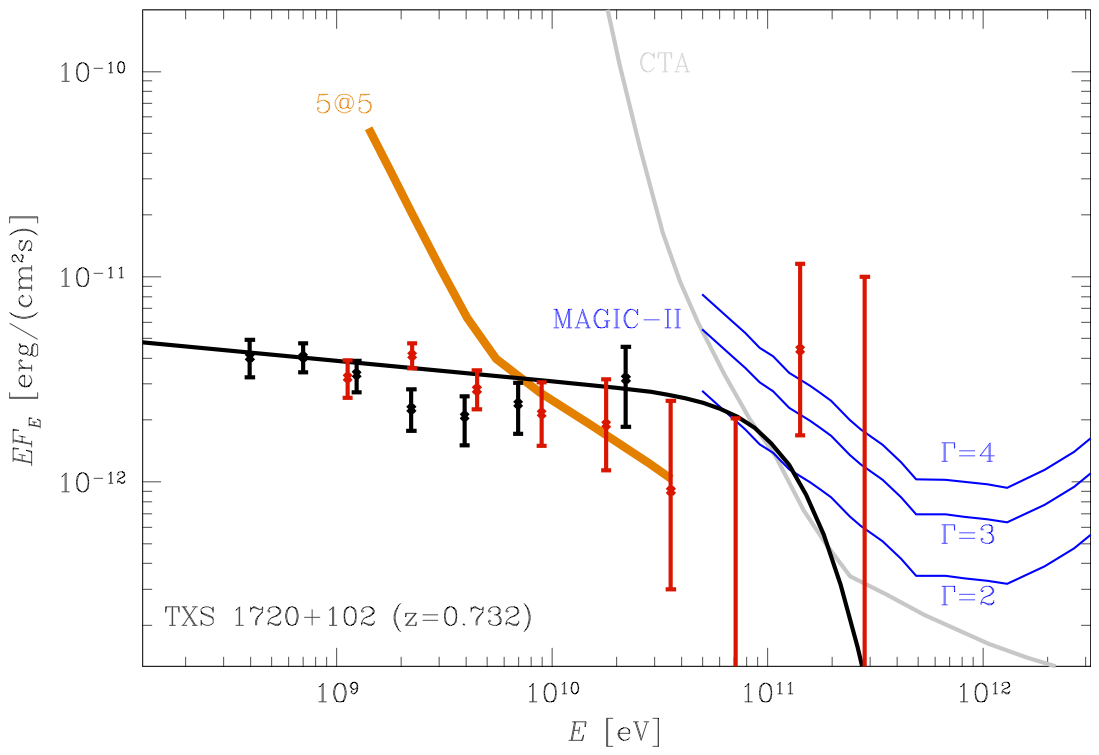}
\caption{Lightcurve (top) and spectrum (bottom) of TXS 1720+102 at the redshift $z=0.732$. Notations are the same as in Figs. \ref{fig:0138} and \ref{fig:0912}.   }
\label{fig:1720}
\end{figure}

This is another representative of the FSRQ sub-population in the high-z VHE source sample. The source is at a redshift $z=0.732$ \citep{afanasiev}. Its spectrum is softer than $\Gamma=2$ in the energy band below 100~GeV. The overall source flux level is somewhat low compared to e.g. {4C +55.17}, so that the source would be hardly detectable by MAGIC, Veritas and/or CTA (see Fig. \ref{fig:1720}), unless the spectrum has a "bump"-like feature in the 100~GeV band. The quality of Fermi data is not sufficient to judge whether this is the case or the Fermi detection in the  $E>100$~GeV is due to the up-fluctuation of the photon count rate. The $E>100$~GeV data are well consistent with the extrapolation of the lower energy powerlaw attenuated the EBL (see Fig. \ref{fig:1720}).

\subsubsection{PKS 1958-179}

\begin{figure}
\includegraphics[width=\linewidth]{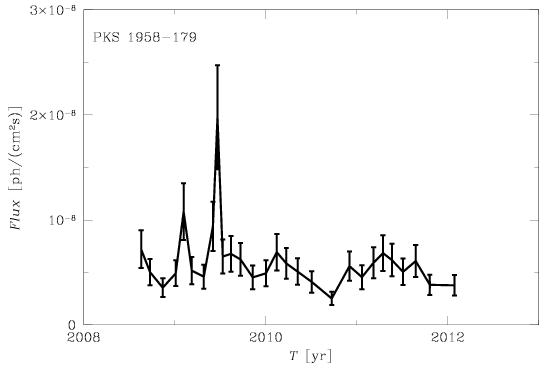}
\includegraphics[width=\linewidth]{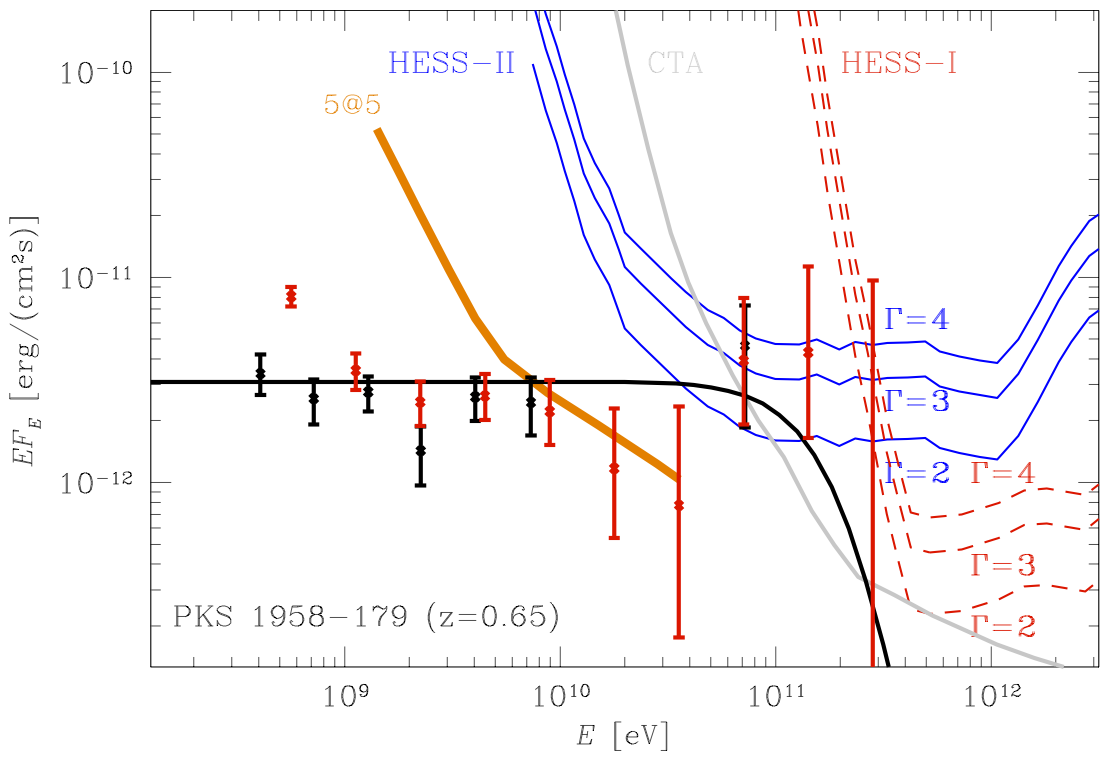}
\caption{Lightcurve (top) and spectrum of PKS1958-179 at the redshift $z=0.65$. Notations are the same as in Fig. \ref{fig:0138}.  }
\label{fig:1958}
\end{figure}

The spectral characteristics of this FSRQ at a redshift of $z=0.65$ \citep{barkhouse} appear somewhat peculiar with a soft spectrum below an energy $\sim 30$~GeV and a possibly a new hard component appearing above the 30~GeV band, see Fig. \ref{fig:1958}. The statistics of the  Fermi data is not sufficient to draw definitive conclusions about the existence of two separate components in the spectrum. In principle, a single powerlaw with a photon index of about $\Gamma=2$ provides a satisfactory fit to the data above an energy of 1~GeV. Within such a spectral model, the detection of the source in the energy band above 100~GeV is consistent with the assumption that the powerlaw extends up to the VHE band, if the absorption of the VHE \gr s through interactions with the EBL photons is taken into account. The source could be marginally detectable with the ground based \gr\ telescopes, including both existing and future facilities (see Fig. \ref{fig:1958}).

\subsubsection{PKS 2142-75}

\begin{figure}
\includegraphics[width=\linewidth]{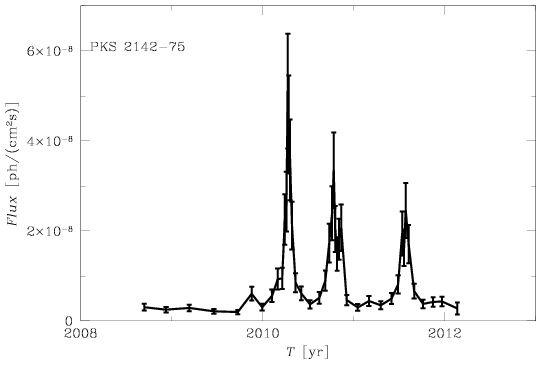}
\includegraphics[width=\linewidth]{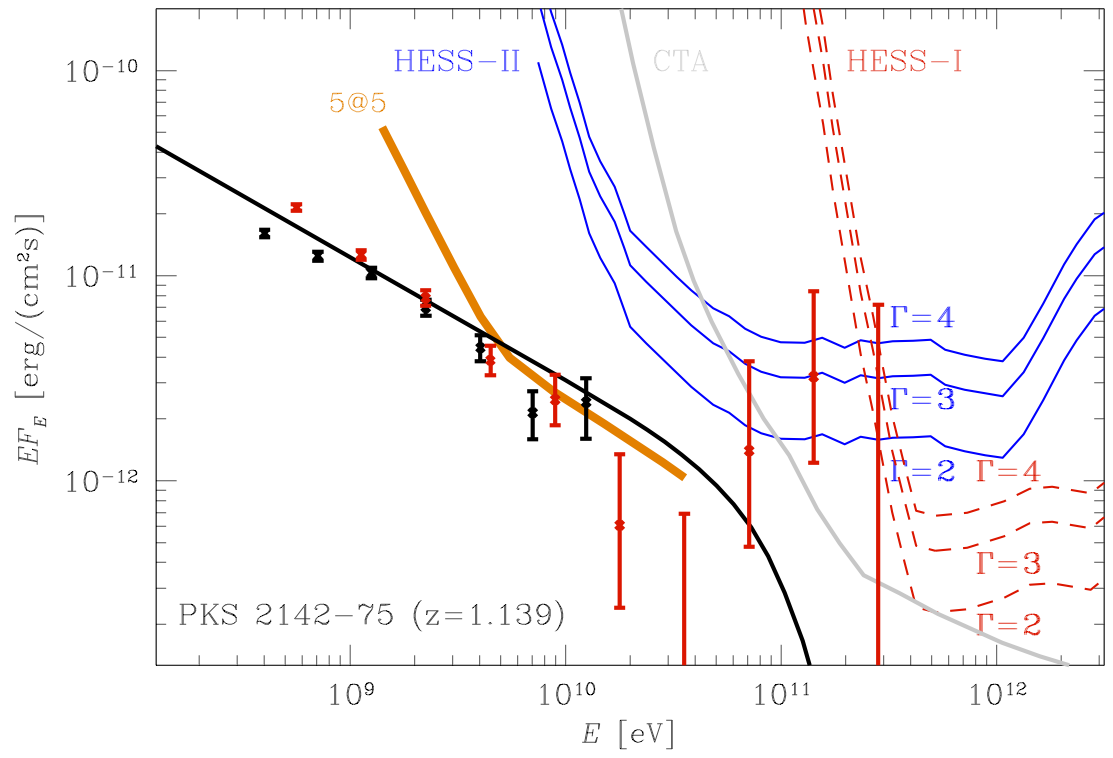}
\caption{Lightcurve (top) and spectrum of PKS 2142-75  at the redshift $z=1.139$. Notations are the same as in Fig. \ref{fig:0138}.   }
\label{fig:2142}
\end{figure}

This FSRQ at $z=1.139$ \citep{jauncey} also has a peculiar time-average spectrum in the 0.3-300~GeV  energy band, with a soft component dominating at energies below 30~GeV and possibly a new hard component appearing above 30~GeV, see Fig. \ref{fig:2142}. If not for this hard component, the source would not be detectable at energies above 100~GeV. During the period of Fermi observations, the source has exhibited several pronounced flares (Fig. \ref{fig:2142} upper panel). The complicated shape of the spectrum might be due to the fact that different components contribute to the source flux during the flare and quiescent periods. However, with only two photons above 30~GeV, one could not make any firm conclusions on the reality and possible nature of the "hard excess". It is well possible that the VHE photon coming from within $0.1^\circ$ of the source is a background photon falling coincident to the source (as discussed in the previous section, one out of the four FSRQs listed in Table \ref{tab:list} could possibly be a false detection in the VHE band).

\subsubsection{RGB J0250+172}

\begin{figure}
\includegraphics[width=\linewidth]{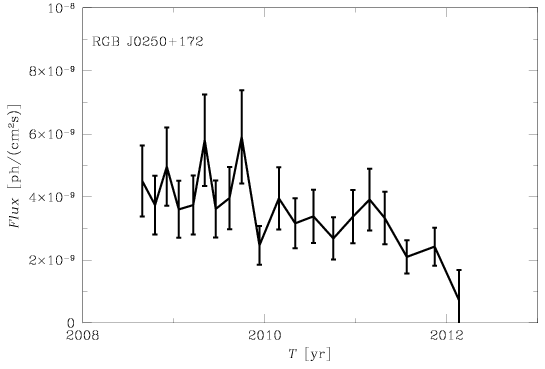}
\includegraphics[width=\linewidth]{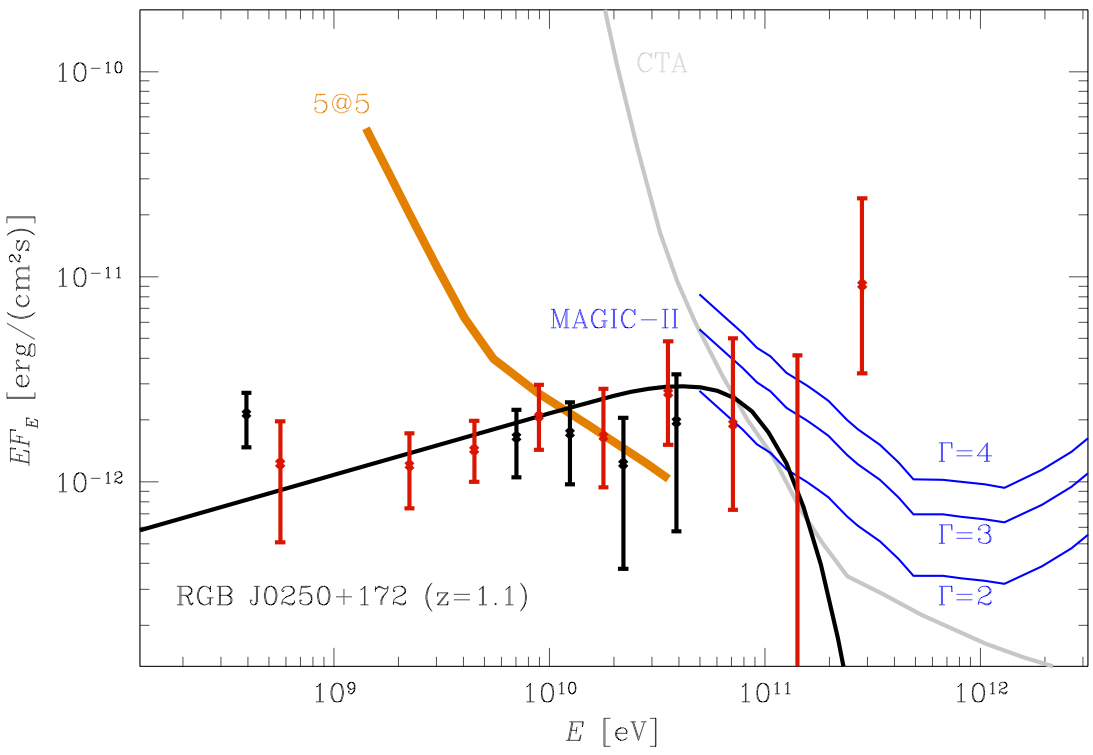}
\caption{Lightcurve (top) and spectrum of RGB J0250+172 at the redshift $z=1.1$.  Notations are the same as in Figs. \ref{fig:0138} and \ref{fig:0912}.  }
\label{fig:0250}
\end{figure}

Contrary to the soft spectra FSRQs,  RGB J0250+172 at $z=1.1$ \citep{bauer} is characterized by a hard \gr\ spectrum and is a good candidate for being a real VHE \gr\ source.  However, due to the lower overall flux normalization, the source should only be marginally detectable by both Fermi and by the current and next-generation ground-based \gr\ telescopes (see Fig. \ref{fig:0250}). Unexpectedly, a photon with energy 358~GeV is detected from the source direction. This detection is somewhat puzzling. On the one side, the estimate of the source flux in the 200-400~GeV band, based on this photon, is consistent with the powerlaw extrapolation of the lower-energy spectrum, without an account of the EBL-induced suppression of the source flux (see Fig. \ref{fig:0250}). On the other side, if the EBL suppression is taken into account, the measured source flux is orders of magnitude above the expected flux from a source at redshift 1.1. Resolution of this inconsistency might be (a) that the VHE photon from the source direction is a background photons, with the chance probability  0.8\%, (b) that the energy of this particular photon is over-estimated by the LAT data analysis software or (c) that the redshift of the source has not been correctly measured. 

This contradiction could be easily resolved via observation of the source with existing ground-based \gr\ telescope(s) like  MAGIC. If the source flux in the 200-400~GeV band is indeed at the level of $\sim 10^{-11}$~erg/(cm$^2$s), it should be readily measurable by MAGIC. If the flux is at the level expected based on the estimate of the model calculation taking into account the EBL suppression, the source is likely not detectable neither by MAGIC nor by CTA. It would be, however, detectable by 5@5 due to the lower energy threshold of the instrument.

\subsubsection{PKS 1130+008}

\begin{figure}
\includegraphics[width=\linewidth]{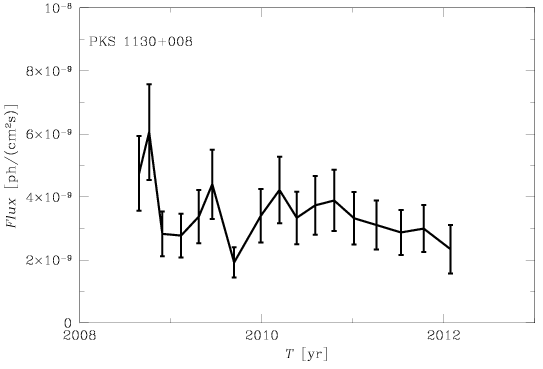}
\includegraphics[width=\linewidth]{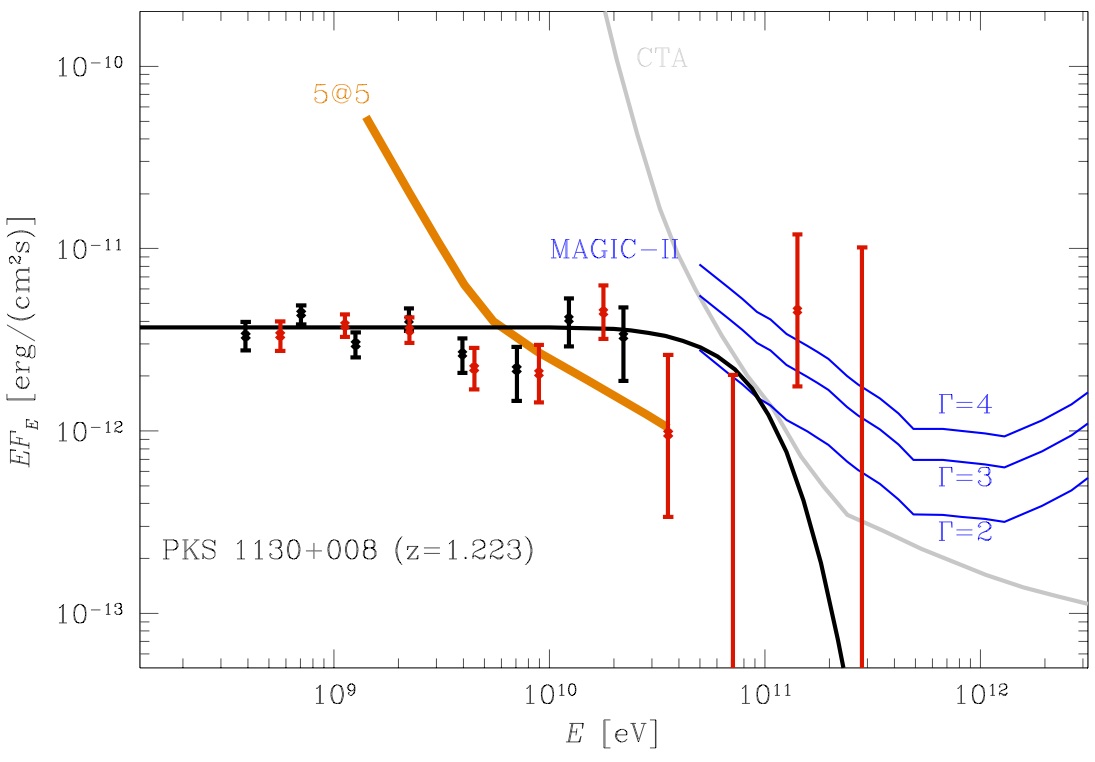}
\caption{Lightcurve (top) and spectrum of PKS 1130+008 at the redshift $z=1.223$.  Notations are the same as in Figs. \ref{fig:0138} and \ref{fig:0912}.}
\end{figure}

The redshift of this source is quoted in SIMBAD to be $z=1.223$ (from Ref. \citep{sdss_0138}. The Fermi AGN catalog \citep{fermi_agn} ascribes the redhsift 0.678 to the source. The source has a powerlaw-type spectrum in the energy range 0.3-30~GeV with photon index $\Gamma\simeq 2$. Significant suppression of the source flux above $\sim 30$~GeV is expected taking into account the high redshift of the source. In this respect, the estimate of the source flux in the 100-200~GeV band, based on the photon detected by the LAT from the source direction, is somewhat surprising, because it lies at the extrapolation of the powerlaw from the lower energies and is above the expected flux level taking into account the expected EBL suppression. At the same time, the inconsistency is still very mild due to the large uncertainty of the Fermi flux estimate. 

A relatively low level of the source flux makes its detection in the $E>100$~GeV band with MAGIC and CTA difficult if not impossible, if the redshift is $z=1.223$ (it is, however, detectable if the redshift is $z=0.678$, as in the Fermi AGN catalog \citep{fermi_agn}). At the same time, a dedicated low-energy-threshold ground-based \gr\ telescope, like 5@5 would be able to study the details of the EBL effects on the source spectrum, because the source would be detectable from an energy $\sim$10~GeV.

\section{Discussion}

The detection of VHE \gr\ emission at energies above 100~GeV from blazars with redshifts beyond $z=0.5$ carries important implications for the possibility of studying the cosmological evolution of the blazar population and the overall energy output from the galaxies, in the form of the EBL. 

\begin{figure}
\includegraphics[height=\linewidth]{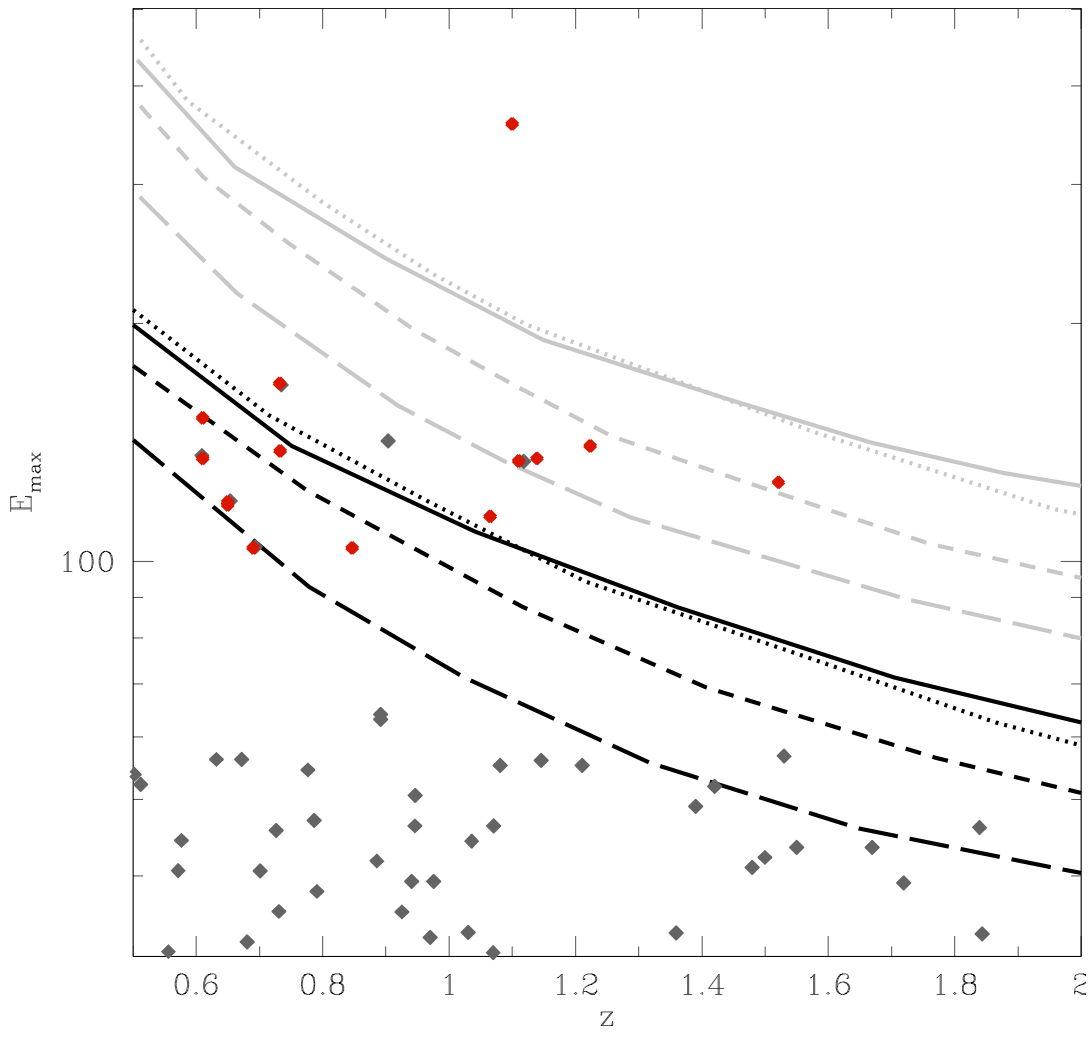}
\caption{Energies of VHE photons from BL Lacs and FSRQ as a function of redshift. Curves correspond to the optical depth $\tau=1$ (black) and $\tau=3$ (grey) with respect to absorption on the EBL. Solid line:  EBL from {\citet{franceschini08}}; dotted line: EBL from {\citet{gilmore}};  short-dashed line: EBL from {\citet{finke}}; long-dashed line: EBL from {\citet{kneiske}}}
\label{fig:photons}
\end{figure}

Fermi, being able to discover the VHE \gr\ signal from the high-redshift sources, does not have a capability of studying the details of the spectral characteristics of these high-redshift sources. At the same time, Fermi's detections of these sources provides a clear indication for the selection of the high-redshift targets for observations with existing and next-generation ground-based \gr\ telescopes. 

In the previous section we have compared the Fermi flux measurements in the $E>100$~GeV band with the sensitivity of the ground-based \gr\ telescopes. The results from such comparisons indicate that the flux levels of sources listed in table \ref{tab:list} are already at or below the sensitivity limit of the current generation ground-based telescopes and are, in fact just at the sensitivity limit of the next generation facility CTA. This means that Table \ref{tab:list} provides a more-or-less exhaustive list of the high-redshift sources accessible for detection using ground-based \gr\ telescopes. 

The situation with the limited capabilities for the detection of high-redshift sources from the ground could, nevertheless, change if a telescope system specially optimized for the reduction of the low-energy threshold, like 5@5 would be realized. In this case most of the sources listed in Table~\ref{tab:list} would be detectable by 5@5 already at energies of about 10~GeV, with very high signal statistics. This would allow a high-quality study of the details of the spectral characteristics of the VHE \gr\ emission from high redshift sources. Lowering the energy threshold with 5@5 would open the possibility for detecting a much larger number of the high-redshift sources, compared to the just thirteen sources listed in Table~\ref{tab:list}. 

Low statistics of the VHE \gr\ signal from the Fermi high redshift blazars does not allow a study of the effect of absorption of \gr s through their interactions with EBL photons  on a source-by-source basis. However, the significance of this effect for different models of cosmological evolution of EBL could be evaluated collectively for all thirteen sources from Table \ref{tab:list}, following the method discussed by \citet{abdo10}. Fig. \ref{fig:photons} shows the energies of VHE \gr s from distant blazars as a function of the source redshift. The data from the previous analysis at lower energies, reported by \citet{abdo10}, are shown in grey in the same figure for comparison. The black/grey curves show the energies at which the optical depth with respect to pair production on the EBL is $\tau=1$ and $\tau=3$, respectively. This implies a flux suppression by a factor of $\simeq 3$ and $\simeq 20$. A naive expectation is that there should be no VHE photons from the sources in the upper part of the plot, much beyond the corresponding $\tau=1$ or $\tau=3$ curves. From Fig. \ref{fig:photons} one can see that this trend of decreasing photon counts beyond the $\tau=1$ curve holds in the redshift range $z=0.5-1$, with only one photon beyond the solid black curve corresponding to $\tau=1$ for the \citet{franceschini08} model of EBL evolution. At the same time, the trend seems to be broken for redshifts of about $z\gtrsim 1$, where a large number of photons beyond the $\tau=1$ curve appears. This large number of photons beyond the $\tau=1$ curve is definitely not due to the larger source number in this redshift range. Table \ref{tab:list} contains seven sources at $z<1$ and six sources at $z\ge 1$.  Thus, the larger statistics of photons beyond the $\tau=1$ curve should be due to a different reason. 

One possibility is that the determination of redshifts of the $z\ge 1$ sources listed in the table \ref{tab:list} are not reliable and, in fact, the sources are at lower redshifts. In the previous section it is mentioned that this might be the case for  B2 0912+29. Another possibility is that  at least one or two photons in the $z>1$ part of the diagram might still be from the background and not from the high-redshift sources. This might be the case for the highest energy photon beyond $\tau=3$ curves, which came from the direction of  RGB J0250+172. This photon has a 0.8\% probability of being from the background. 

If none of these possible explanations holds (this needs to be checked with more data and verification of the redshift measurements), then the over-abundance of the VHE photons beyond the $\tau=1$ curves in the redshift range $z>1$ in Fig. \ref{fig:photons} should be due a physical effect. One possibility is that  the EBL level at high redshifts is lower, so that the Universe is more transparent to the VHE photons, than what is implied by the currently existing models.  Otherwise, new propagation effects, such as  origin of the \gr\ emission in the ultra-high-energy cosmic ray induced cascade \citep{Essey2009,murase,kusenko}
or even by new physics, like axions \citep{axions}, should be considered.    

In the scenario of \citet{Essey2009}, UHE protons  with energies in the EeV range propagate cosmological distances and lose energy primarily through the proton pair production process. 
In this case secondary TeV gamma-rays are produced by such protons at distances 100-300 Mpc and easily reach the Earth.  For this model to be valid,  one needs a relatively low extra-galactic magnetic field with values 1-10 pG everywhere on the way of the protons, not only in the voids of large scale structure. Apparently, a support for such a scenario is found in a recent work of \citet{Landt2012}, in which absorption lines at the redshift beyond $z=1$ were found in the spectra of two BL Lacs, PKS 0447-439 and PMN J0630-24. One of these BL Lacs has been recently observed in the VHE band by HESS {\citep{zech}. Explanation of the detection of PKS 0447-439 in the VHE band within conventional models would be challenging if the source redshift is indeed $z>1$  \citep{kusenko}. Thorough verification of the redshift measurement for both PKS 0447-439 and for the $z\ge 1$ sources listed in the Table \ref{tab:list} is extremely important for the clarificaiton of consistency / inconsistency of the VHE band detections of the high-redshift sources with the current understanding of the mechanisms of formation of the VHE \gr\ spectra of the sources and of the cosmological evolution of the EBL. 

\section*{Acklowledgements}

The work of AN, AT and IeV is supported by the Swiss Naitonal Science Foundation grant PP00P2\_123426. AN acknowledges the hospitality of Paris Astronomical Observatory in Meudon where this work was completed.

\end{document}